\documentstyle[prl,aps,floats,epsf,color]{revtex}
\begin{document}
\baselineskip=12pt
\def\be{\begin{equation}}
\def\ee{\end{equation}}
\def\bea{\begin{eqnarray}}
\def\eea{\end{eqnarray}}
\def\E{{\rm e}}
\def\bearst{\begin{eqnarray*}}
\def\eearst{\end{eqnarray*}}
\def\peleven{\parbox{11cm}}
\def\peffec{\peight{\bearst\eearst}\hfill\peleven}
\def\pspace{\peight{\bearst\eearst}\hfill}
\def\ptwelve{\parbox{12cm}}
\def\peight{\parbox{8mm}}
\twocolumn[\hsize\textwidth\columnwidth\hsize\csname@twocolumnfalse\endcsname

\title
{Multifractal Detrended Fluctuation Analysis of Sunspot Time
Series }
\author{M. Sadegh Movahed$^{1,2,3}$, G. R. Jafari$^{2,4}$, F. Ghasemi$^{2}$, Sohrab Rahvar$^{1,2}$
and M. Reza Rahimi Tabar$^{1,5}$, }

\address{$^{1}$Department of Physics, Sharif University of
Technology, P.O.Box 11365--9161, Tehran, Iran}
\address{$^{2}$ Institute for Studies in theoretical Physics and Mathematics, P.O.Box 19395-5531,Tehran, Iran}
\address{$^{3}$Iran Space Agency,  PO Box  199799-4313, Tehran,
Iran}
\address{$^{4}$Department of Physics, Shahid Beheshti University, Evin, Tehran 19839, Iran}
\address{$^{5}$CNRS
UMR 6529, Observatoire de la C$\hat o$te d'Azur, BP 4229, 06304
Nice Cedex 4, France}

\vskip 1cm

 \maketitle
%\date{00/07/2000}
%\maketitle

%%%%%%%%%%%%%%%%%%%%%%%%%%%%%%%%%%%%%%%%%%%%%%%%%%%%%%
%ABSTRACT
%%%%%%%%%%%%%%%%%%%%%%%%%%%%%%%%%%%%%%%%%%%%%%%%%%%%%%

\begin{abstract}

 We use multifractal detrended fluctuation analysis (MF-DFA), to See
query 1 study sunspot number fluctuations. The result of the
MF-DFA shows that there are three crossover timescales in the
fluctuation function. We discuss how the existence of the
crossover timescales is related to a sinusoidal trend. Using
Fourier detrended fluctuation analysis, the sinusoidal trend is
eliminated. The Hurst exponent of the time series without the
sinusoidal trend is $0.12\pm 0.01$. Also we find that these
fluctuations have multifractal nature. Comparing the MF-DFA
results for the remaining data set to those for shuffled and
surrogate series, we conclude that its multifractal nature is
almost entirely due to long range correlations.\\
Keyboard: New applications of statistical mechanics

\end{abstract}

\hspace{.3in}
\newpage
 ]
\section{Introduction}
\begin{figure}
\epsfxsize=8truecm\epsfbox{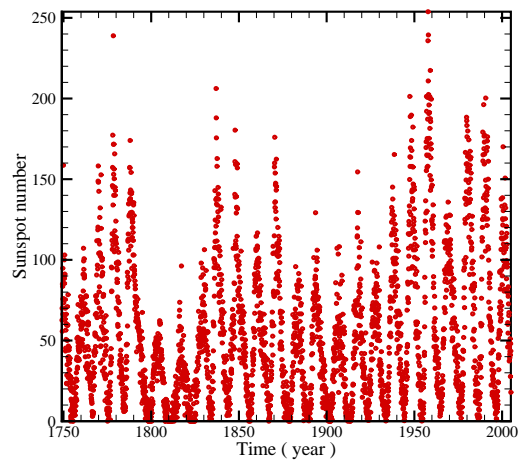} \narrowtext \caption{Observed
spot numbers as a function of time.} \label{fig1}
 \end{figure}

The important feature of the sun`s outer regions is the existence
of a reasonably strong magnetic field. To the lowest order of
approximation, the sun's magnetic field is dipolar in character
and is axisymmetric. The strength of the field on a typical point
on the solar surface is approximately a few Gauss. There is,
however significant variation in this value and there are
localized regions (called $sunspots$) in which the filed can be
much higher \cite{bray}. Because of the symmetry of the twisted
magnetic lines as the origin of sunspots, they are generally seen
in pairs or in groups of pairs at both sides of the solar equator.
As the sunspot cycle progresses, spots appear closer to the sun's
equator giving rise to the so called ``butterfly diagram'' in the
time latitude distribution \cite{petr00}. The twisted magnetic
field above sunspots are sites where solar flares are observed. It
has been found that chromospheric flares show a very close
statistical relationship with sunspots \cite{bray}. The number of
sunspots is continuously changing in time in a random fashion and
constitutes a typically random time series. Figure \ref{fig1}
shows the monthly measured number of sunspots in terms of time.
The data belongs to a data set collected by the Sunspot Index Data
Center (SIDC) from 1749 up to now \cite{data}.

Recently, the statistical properties of sun activity have been
investigated by some methods in chaos theory \cite{veronig02} and
multifractal analysis \cite{valentyana05,zhukov03}. The periodical
occurrence of hemispheric sunspot numbers have been analyzed with
respect to the changes in time using wavelet. The north-south
asymmetries concerning solar activity and rotational behavior has
been investigated by using the wavelet and auto-correlation
function \cite{ter002}. Cross-correlation functions between
monthly mean sunspot areas and sunspot numbers have been
determines in some papers \cite{tem02}. The evidence for the
existence of ''active longitudes'' on the sun is given by using
the autocorrelation function of daily sunspot numbers
\cite{tem02,bog82}. See also \cite{ter02,ter03} about the relation
between sunspot number fluctuation and number of flares, their
evolution step, i.e. duration, rise times, decay times, event
asymmetries.

%%%%%%%%%%%%%%%%%%%%%%%%%%%%%%%%%%%%%%%%%%%%%%%%%%%%%%%%%%%%%%%%%%%%%%%%%
%%%%%%%%%%%%%%%%%%%%%%%%%%%%%%%%%%%%%%%%%%%%%%%%%%%%%%%%%%%%%%%%%

In this paper we would like to characterize the complex behavior
of sunspot time series through the computation of the signal
parameters - scaling exponents - which quantifies the correlation
exponents and multifractality of the signal. As shown in Figure
\ref{fig1}, the sunspot time series has a sinusoidal trend, with a
frequency is equal to the well known cycle of sun activity,
approximately $11$years. Because of the nonstationary nature of
sunspot time series, and due to the finiteness of the available
data sample, we should apply some methods which are insensitive to
non-stationarities like trends.

To eliminate the effect of sinusoidal trend, we apply the Fourier
Detrended Fluctuation Analysis (F-DFA) \cite{na04,chi05}. After
elimination of the trend we use the Multifractal Detrended
Fluctuation Analysis (MF-DFA) to analysis the data set. The MF-DFA
 methods are the modified version of detrended fluctuation
analysis (DFA) to detect multifractal properties of time series.
The detrended fluctuation analysis (DFA) method introduced by Peng
et al. \cite{Peng94} has became a widely-used technique for the
determination of (mono-) fractal scaling properties and the
detection of long-range correlations in noisy, nonstationary time
series \cite{Peng94,murad,physa,kunhu,kunhu1}. It has successfully
been applied to diverse fields such as DNA sequences
\cite{Peng94,dns}, heart rate dynamics \cite{herz,Peng95,PRL00},
neuron spiking \cite{neuron}, human gait \cite{gait}, long-time
weather records \cite{wetter}, cloud structure \cite{cloud},
geology \cite{malamudjstatlaninfer1999}, ethnology
\cite{Alados2000}, economical time series \cite{economics}, and
solid state physics \cite{fest}.

The paper is organized as follows:  In Section II we describe the
MF-DFA and F-DFA methods in detail and show that the scaling
exponents determined via the MF-DFA method are identical to those
obtained by the standard multifractal formalism based on partition
functions. We eliminate the sinusoidal trend via the F-DFA
technique in Section III and investigate the multifractal nature
of the remaining fluctuation. In Section IV, we examine the source
of multifractality in sunspot data by comparison the MF-DFA
results for remaining data set to those obtained
via the MF-DFA for shuffled and surrogate series. Section V closes with a discussion of the present results.\\
\section{Multifractal Detrended Fluctuation Analysis}

The simplest type of the multifractal analysis is based upon the
standard partition function multifractal formalism, which has been
developed for the multifractal characterization of normalized,
stationary measurements \cite{feder88,barabasi,peitgen,bacry01}.
Unfortunately, this standard formalism does not give correct
results for nonstationary time series that are affected by trends
or that cannot be normalized. Thus, in the early 1990s an improved
multifractal formalism has been developed, the wavelet transform
modulus maxima (WTMM) method \cite{wtmm}, which is based on the
wavelet analysis and involves tracing the maxima lines in the
continuous wavelet transform over all scales. The other method,
the multifractal detrended fluctuation analysis (MF-DFA), is based
on the identification of scaling of the $q$th-order moments
depending on the signal length and is generalization of the
standard DFA using only the second moment $q=2$.

The MF-DFA does not require the modulus maxima procedure in
contrast WTMM method, and hence does not require more effort in
programming and computing than the conventional DFA. On the other
hand, often experimental data are affected by non-stationarities
like trends, which have to be well distinguished from the
intrinsic fluctuations of the system in order to find the correct
scaling behavior of the fluctuations. In addition very often we do
not know the reasons for underlying trends in collected data and
even worse we do not know the scales of the underlying trends,
also, usually the available record data is small. For the reliable
detection of correlations, it is essential to distinguish trends
from the fluctuations intrinsic in the data. Hurst rescaled-range
analysis \cite{hurst65} and other non-detrending methods work well
if the records are long and do not involve trends. But if trends
are present in the data, they might give wrong results. Detrended
fluctuation analysis (DFA) is a well-established method for
determining the scaling behavior of noisy data in the presence of
trends without knowing their origin and shape
\cite{Peng94,Peng95,fano,allan,buldy95}

%This task is not easy, since e.~g. subtracting some kind of
%moving average with a certain bin width $\sigma$ would
%artificially introduce the time scale $\sigma$ into the data,
%thus destroying a possible scaling over a wider range of time
%scales. .  . .
%***********************************************************************
%One reason to employ the DFA method is to avoid spurious
%detection of correlations that are artefacts of nonstationarity
%in time series.
%*********************************************************************

\subsection{Description of the MF-DFA}

The modified multifractal DFA (MF-DFA) procedure consists of five
steps.  The first three steps are essentially identical to the
conventional DFA procedure (see e.~g.
\cite{Peng94,murad,physa,kunhu,kunhu1}). Suppose that $x_k$ is a
series of length $N$, and that this series is of compact support,
i.e. $x_k = 0$ for an insignificant fraction of the values only.

\noindent $\bullet$ {\it Step 1}: Determine the ``profile''
\begin{equation} Y(i) \equiv \sum_{k=1}^i \left[ x_k - \langle x
\rangle \right], \qquad i=1,\ldots,N. \label{profile}
\end{equation}
Subtraction of the mean $\langle x \rangle$ is not compulsory,
since it would be eliminated by the later detrending in the third
step.

\noindent $\bullet$ {\it Step 2}: Divide the profile $Y(i)$ into
$N_s \equiv {\rm int}(N/s)$ non-overlapping segments of equal
lengths $s$. Since the length $N$ of the series is often not a
multiple of the considered time scale $s$, a short part at the end
of the profile may remain.  In order not to disregard this part of
the series, the same procedure is repeated starting from the
opposite end.  Thereby, $2 N_s$ segments are obtained altogether.

\noindent $\bullet$ {\it Step 3}: Calculate the local trend for
each of the $2 N_s$ segments by a least-square fit of the series.
Then determine the variance
\begin{equation} F^2(s,\nu) \equiv {1 \over s} \sum_{i=1}^{s}
\left\{ Y[(\nu-1) s + i] - y_{\nu}(i) \right\}^2, \label{fsdef}
\end{equation}
for each segment $\nu$, $\nu = 1, \ldots, N_s$ and
\begin{equation} F^2(s,\nu) \equiv {1 \over s} \sum_{i=1}^{s}
\left\{ Y[N - (\nu-N_s) s + i] - y_{\nu}(i) \right\}^2,
\label{fsdef2}
\end{equation}
for $\nu = N_s+1, \ldots, 2 N_s$.  Here, $y_{\nu}(i)$ is the
fitting polynomial in segment $\nu$.  Linear, quadratic, cubic, or
higher order polynomials can be used in the fitting procedure
(conventionally called DFA1, DFA2, DFA3, $\ldots$)
\cite{Peng94,PRL00}. Since the detrending of the time series is
done by the subtraction of the polynomial fits from the profile,
different order DFA differ in their capability of eliminating
trends in the series.  In (MF-)DFA$m$ [$m$th order (MF-)DFA]
trends of order $m$ in the profile (or, equivalently, of order $m
- 1$ in the original series) are eliminated.  Thus a comparison of
the results for different orders of DFA allows one to estimate the
type of the polynomial trend in the time series
\cite{physa,kunhu}.

\noindent $\bullet$ {\it Step 4}: Average over all segments to
obtain the $q$-th order fluctuation function, defined as:
\begin{equation} F_q(s) \equiv \left\{ {1 \over 2 N_s}
\sum_{\nu=1}^{2 N_s} \left[ F^2(s,\nu) \right]^{q/2}
\right\}^{1/q}, \label{fdef}\end{equation}
where, in general, the index variable $q$ can take any real value
except zero.  For $q=2$, the standard DFA procedure is retrieved.
Generally we are interested in how the generalized $q$ dependent
fluctuation functions $F_q(s)$ depend on the time scale $s$ for
different values of $q$.  Hence, we must repeat steps 2, 3 and 4
for several time scales $s$.  It is apparent that $F_q(s)$ will
increase with increasing $s$.  Of course, $F_q(s)$ depends on the
DFA order $m$. By construction, $F_q(s)$ is only defined for $s
\ge m+2$.

\noindent $\bullet$ {\it Step 5}: Determine the scaling behavior
of the fluctuation functions by analyzing log-log plots of
$F_q(s)$ versus $s$ for each value of $q$. If the series $x_i$ are
long-range power-law correlated, $F_q(s)$ increases, for large
values of $s$, as a power-law,
\begin{equation} F_q(s) \sim s^{h(q)} \label{Hq}. \end{equation}
In general, the exponent $h(q)$ may depend on $q$.   For
stationary time series such as  fGn (fractional Gaussian noise),
$Y(i)$ in Eq. \ref{profile}, will be a fBm (fractional Brownian
motion) signal, so, $0<h(q=2)<1.0$. The exponent $h(2)$ is
identical to the well-known Hurst exponent $H$
\cite{Peng94,murad,feder88}. Also for a nonstationary  signal,
such as fBm noise, $Y(i)$ in Eq. \ref{profile}, will be a sum of
fBm signal, so the corresponding scaling exponent of $F_q(s)$ is
identified by $h(q=2)>1.0$ \cite{Peng94,eke02} (see the appendix
for more details). In this case the relation between the exponents
$h(2)$ and $H$ will be $H=h(q=2)-1$. The exponent $h(q)$ is known
as generalized Hurst exponent.  The auto-correlation function can
be characterized by a power law $C(s)\equiv\langle n_kn_{k+s}
\rangle \sim s^{-\gamma}$ with exponent $\gamma=2-2H$. Its power
spectra can be characterized by $S(\omega)\sim\omega^{-\beta}$
with frequency $\omega$ and $\beta=2H-1$, In the nonstationary
case, correlation exponent and power spectrum scaling are
$\gamma=-2H$ and $\beta=2H+1$, respectively \cite{Peng94,eke02}.

For monofractal time series, $h(q)$ is independent of $q$, since
the scaling behavior of the variances $F^2(s,\nu)$ is identical
for all segments $\nu$, and the averaging procedure in
Eq.~(\ref{fdef}) will just give this identical scaling behavior
for all values of $q$. If we consider positive values of $q$, the
segments $\nu$ with large variance $F^2(s,\nu)$ (i.~e. large
deviations from the corresponding fit) will dominate the average
$F_q(s)$.  Thus, for positive values of $q$, $h(q)$ describes the
scaling behavior of the segments with large fluctuations. For
negative values of $q$, the segments $\nu$ with small variance
$F^2(s,\nu)$ will dominate the average $F_q(s)$. Hence, for
negative values of $q$, $h(q)$ describes the scaling behavior of
the segments with small fluctuations \cite{note3}.

\subsection{Relation to standard multifractal analysis}
%*************************************************************

%***************************************************************
For a stationary, normalized series the multifractal scaling
exponents $h(q)$ defined in Eq.~(\ref{Hq}) are directly related to
the scaling exponents $\tau(q)$ defined by the standard partition
function-based multifractal formalism as shown below. Suppose that
the series $x_k$ of length $N$ is a stationary, normalized
sequence. Then the detrending procedure in step 3 of the MF-DFA
method is not required, since no trend has to be eliminated. Thus,
the DFA can be replaced by the standard Fluctuation Analysis (FA),
which is identical to the DFA except for a simplified definition
of the variance for each segment $\nu$, $\nu = 1, \ldots, N_s$.
Step 3 now becomes [see Eq.~(\ref{fsdef})]:
\begin{equation} F_{\rm FA}^2(s,\nu) \equiv [Y(\nu s) - Y((\nu-1) s)]^2.
\label{FAfsdef} \end{equation}
Inserting this simplified definition into Eq.~(\ref{fdef}) and
using Eq.~(\ref{Hq}), we obtain
\begin{equation} \left\{ {1 \over 2 N_s} \sum_{\nu=1}^{2 N_s}
\vert Y(\nu s) - Y((\nu-1) s) \vert^q \right\}^{1/q} \sim
s^{h(q)}. \label{FAfHq} \end{equation}
For simplicity we can assume that the length $N$ of the series is
an integer multiple of the scale $s$, obtaining $N_s = N/s$ and
therefore
\begin{equation} \sum_{\nu=1}^{N/s} \vert Y(\nu s) - Y((\nu-1) s)
\vert^q \sim s^{q h(q) - 1}. \label{MFA} \end{equation}
This corresponds to the multifractal formalism used e.~g. in
\cite{barabasi,bacry01}. In fact, a hierarchy of exponents $H_q$
similar to our $h(q)$ has been introduced based on Eq.~(\ref{MFA})
in \cite{barabasi}. In order to relate also to the standard
textbook box counting formalism \cite{feder88,peitgen}, we employ
the definition of the profile in Eq.~(\ref{profile}). It is
evident that the term $Y(\nu s) - Y((\nu-1) s)$ in Eq.~(\ref{MFA})
is identical to the sum of the numbers $x_k$ within each segment
$\nu$ of size $s$. This sum is known as the box probability
$p_s(\nu)$ in the standard multifractal formalism for normalized
series $x_k$,
\begin{equation} p_s(\nu) \equiv \sum_{k=(\nu-1) s +1}^{\nu s} x_k =
Y(\nu s) - Y((\nu-1) s).  \label{boxprob} \end{equation}
The scaling exponent $\tau(q)$ is usually defined via the
partition function $Z_q(s)$,
\begin{equation} Z_q(s) \equiv \sum_{\nu=1}^{N/s} \vert p_s(\nu)
\vert^q \sim s^{\tau(q)}, \label{Zq} \end{equation}
where $q$ is a real parameter as in the MF-DFA method, discussed
above. Using Eq.~(\ref{boxprob}) we see that Eq.~(\ref{Zq}) is
identical to Eq.~(\ref{MFA}), and obtain analytically the relation
between the two sets of multifractal scaling exponents,
\begin{equation} \tau(q) = q h(q) - 1. \label{tauH} \end{equation}
Thus, we observe that $h(q)$ defined in Eq.~(\ref{Hq}) for the
MF-DFA is directly related to the classical multifractal scaling
exponents $\tau(q)$.  Note that $h(q)$ is different from the
generalized multifractal dimensions
\begin{equation} D(q) \equiv {\tau(q) \over q-1} =
{q h(q)-1 \over q-1}, \label{Dq} \end{equation} that are used
instead of $\tau(q)$ in some papers.  While $h(q)$ is independent
of $q$ for a monofractal time series, $D(q)$ depends on $q$ in
this case. Another way to characterize a multifractal series is
the singularity spectrum $f(\alpha)$, that is related to $\tau(q)$
via a Legendre transform \cite{feder88,peitgen},
\begin{equation} \alpha = \tau'(q) \quad {\rm and} \quad
f(\alpha) = q \alpha - \tau(q). \label{Legendre} \end{equation}
Here, $\alpha$ is the singularity strength or H\"older exponent,
while $f(\alpha)$ denotes the dimension of the subset of the
series that is characterized by $\alpha$. Using Eq.~(\ref{tauH}),
we can directly relate $\alpha$ and  $f(\alpha)$ to $h(q)$,
\begin{equation} \alpha = h(q) + q h'(q) \quad {\rm and} \quad
f(\alpha) = q [\alpha - h(q)] + 1.\label{Legendre2} \end{equation}

A H\"older exponent denotes monofractality, while in the
multifractal case, the different parts of the structure are
characterized by different values of $\alpha$, leading to the
existence of the spectrum $f(\alpha)$.

\subsection{Fourier-Detrended Fluctuation Analysis}

In some cases, there exist one or more crossover (time) scales
$s_\times$ separating regimes with different scaling exponents
\cite{physa,kunhu}. In this case investigation of the scaling
behavior is more complicate and different scaling exponents are
required for different parts of the series \cite{kunhu1}.
Therefore one needs a multitude of scaling exponents
(multifractality) for a full description of the scaling behavior.
A crossover usually can arise from a change in the correlation
properties of the signal at different time or space scales, or can
often arise from trends in the data. To remove the crossover due
to a trend such as sinusoidal trends, Fourier-Detrende Fluctuation
Analysis (F-DFA) is applied.  The F-DFA is a modified approach for
the analysis of low frequency trends added to a noise in time
series \cite{na04,chi05,koscielny98,koscielny98b}.

In order to investigate how we can remove trends having a low
frequency periodic behavior, we transform data record to Fourier
space, then we truncate the first few coefficient of the Fourier
expansion and inverse Fourier transform the series. After removing
the sinusoidal trends we can obtain the fluctuation exponent by
using the direct calculation of the MF-DFA. If truncation numbers
are sufficient, The crossover due to a sinusoidal trend in the
log-log plot of $F_q(s)$ versus $s$ disappears.

\section{Analysis of sunspot time series} \label{detcross}

\begin{figure}
\epsfxsize=8truecm\epsfbox{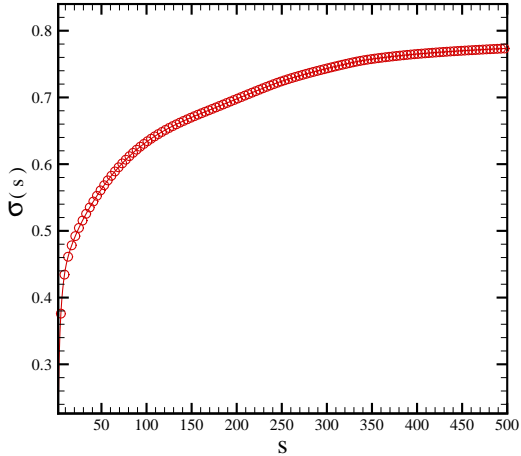} \narrowtext \caption{Behavior
of standard deviation of sunspot time series as a function of time
scale. It shows that this time series is not stationary and direct
calculation of correlation gives strongly wrong result.}
\label{fig2}
 \end{figure}

\begin{figure}
\epsfxsize=8truecm\epsfbox{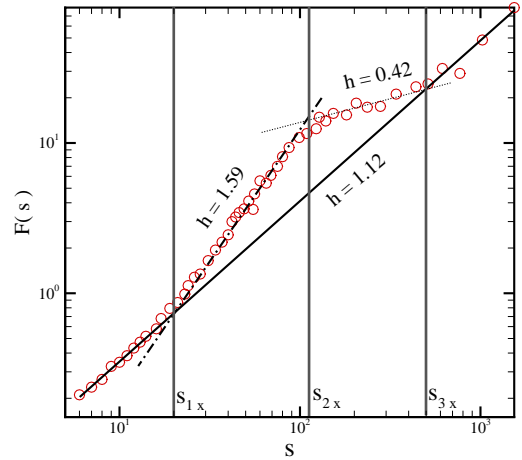} \narrowtext
\caption{Crossover behavior of log-log plot $F(s)$ versus $s$ for
sunspot time series  for $q=2.0$. There are three crossover time
scales in plot of $F(s)$, at scales $s_{1\times}$, $s_{2\times}$
and $s_{3\times}$.} \label{fig3}
 \end{figure}

As mentioned in section II, a spurious of correlations may be
detected if time series is nonstationarity, so direct calculation
of correlation behavior, spectral density exponent, fractal
dimensions etc., don't give the reliable results. It can be
checked that the sunspot time series is nonstationary. One can
verified the non-stationarity property experimentally by measuring
the stability of the average and variance in a moving window for
example with scale $s$. Figure~\ref{fig2} shows the standard
deviation of signal verses scale $s$ isn't saturate. Let us
determine that whether the data set has a sinusoidal trend or not.
According to the MF-DFA1 method, Generalized Hurst exponents
$h(q)$ in Eq.~(\ref{Hq}) can be found by analyzing log-log plots
of $F_q(s)$ versus $s$ for each $q$. Our investigation shows that
there are three crossover time scales $s_{\times}$ in the log-log
plots of $F_q(s)$ versus $s$ for every $q$'s. These three
crossovers divide $F_q(s)$ into four regions, as shown in
Figure~\ref{fig3} ( for instance we took $q=2$). The existence of
these regions is due to the competition between noise and
sinusoidal trend. For $s<s_{1\times}$ and $s>s_{3\times}$, the
noise has the dominating effect \cite{kunhu}. For
$s_{1\times}<s<s_{2\times}$ and $s_{2\times}<s<s_{3\times}$, the
sinusoidal trend dominates \cite{kunhu}. The value of
$s_{2\times}$ is approximately equal to $130$ month which is equal
to the well known cycle of sun activity. As mentioned before, for
very small scales $s<s_{1\times}$ the effect of the sinusoidal
trend is not pronounced, indicating that in this scale region the
signal can be considered as noise fluctuating around a constant
which is filtered out by the MF-DFA1 procedure. In this region the
generalized DFA1 exponent is $h(q=2)=1.12\pm0.01$, where confirms
that the process is a non-stationary process with anti-correlation
behavior.
%%%%%%%%%%%%%%%%%%%%%%%%%%%%%%%%%%%%%%%%%%%%%%%%%%%%%%%%%%%%%%%%%%%%%%%%%
%%%%%%%%%%%%%%%%%%%%%%%%%%%%%%%%%%%%%%%%%%%%%%%%%%%%%%%%%%%%%%%%%

\begin{figure}
\epsfxsize=8truecm\epsfbox{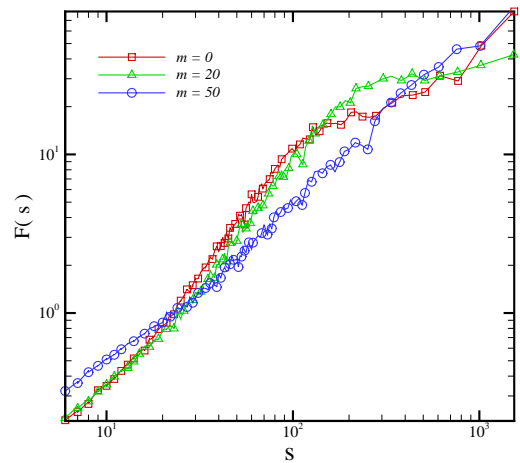} \narrowtext \caption{The
MF-DFA1 functions $F_q(s)$ for sunspot time series versus the time
scale $s$ in log-log plot. Original time series $m=0$, truncation
of the first $20$ terms $m=20$ and $50$ terms $m=50$.}
\label{fig4}
 \end{figure}

 To cancel the sinusoidal trend in
MF-DFA1, we apply F-DFA method for sunspot data. We truncate some
of the first coefficient of the Fourier expansion of sunspot
series. According to Figure \ref{fig4}, for eliminating the
crossover scales, we need to remove approximately $50$ terms of
the Fourier expansion. Then, by inverse Fourier Transformation,
the noise without sinusoidal trend is extracted.

\begin{figure}
\epsfxsize=8truecm\epsfbox{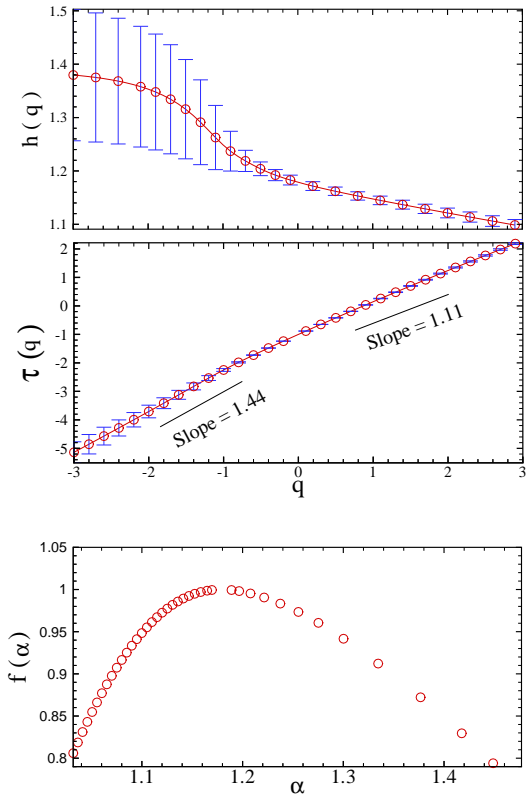} \narrowtext \caption{The $q$
dependence of the generalized Hurst exponent $h(q)$, the
corresponding $\tau(q)$ and singularity spectrum $f(\alpha)$  are
shown in the upper to lower panel respectively for sunspot time
series without sinusoidal trend.} \label{fig5}
 \end{figure}

The MF-DFA1 results of the remanning new signal are shown in
Figure \ref{fig5}. The sunspot time series is a multifractal
process as indicated by the strong $q$ dependence of generalized
Hurst exponents and $\tau(q)$\cite{bun02}. The $q$- dependence of
the classical multifractal scaling exponent $\tau(q)$ has
different behaviors for $q<0$ and $q>0$. For positive and negative
values of $q$, the slopes of $\tau(q)$ are $1.11\pm0.01$ and
$1.44\pm0.01$, respectively. According to the relation between the
Hurst exponent and $h(2)$, i.e. $h(q=2)-1=H$, we find that the
Hurst exponent is $0.12\pm0.01$. This result is equal to value of
Hurst exponent in small scale of MF-DFA1 of noise with sinusoidal
trend. The fractal dimension is obtained as $ D_f = 2 - H = 1.88$
\cite{physa}. The values of derived quantities from MF-DFA1
method, are given in Table \ref{Tab1} and Table \ref{Tab2}.
%%%%%%%%%%%%%%%%%%%%%%%%%%%%%%%%%%%%%%%%%%%%%%%%%%%%%%%%%%%%%%%%%%%%%%%%%
%%%%%%%%%%%%%%%%%%%%%%%%%%%%%%%%%%%%%%%%%%%%%%%%%%%%%%%%%%%%%%%%%

%%%%%%%%%%%%%%%%%%%%%%%%%%%%%%%%%%%%%%%%%%%%%%%%%%%%%%%%%%%%%%%%%%%%%%%%%
%%%%%%%%%%%%%%%%%%%%%%%%%%%%%%%%%%%%%%%%%%%%%%%%%%%%%%%%%%%%%%%%%

Usually, in the MF-DFA method, deviation from a straight line in
the log-log plot of Eq.~(\ref{Hq}) occurs for small scales $s$.
This deviation limits the capability of DFA to determine the
correct correlation behavior for very short scales and in the
regime of small $s$. The modified MF-DFA is defined as follows
\cite{physa}:

\begin{eqnarray} F^{\rm mod}_q(s) &=& \frac{F_{q}(s)}{K_{q}(s)},\nonumber\\
& =& F_q(s) {\langle [F_q^{\rm shuf}(s')]^2 \rangle^{1/2} \,
s^{1/2} \over \langle [F_q^{\rm shuf}(s)]^2 \rangle^{1/2} \,
s'^{1/2} } \quad {\rm (for} \, s' \gg 1),\nonumber\\
\label{fmod}\end{eqnarray}
 where $\langle [F_q^{\rm shuf}(s)]^2
\rangle^{1/2}$ denotes the usual MF-DFA fluctuation function
[defined in Eq.~(\ref{fdef})] averaged over several configurations
of shuffled data taken from the original time series, and $s'
\approx N/40$. The values of the Hurst exponent obtained by
modified MF-DFA1 methods for sunspot time series is $0.11\pm0.01$.
The relative deviation of the Hurst exponent which is obtained by
modified MF-DFA1 in comparison to MF-DFA1 for original data is
approximately $8.33\%$.
%%%%%%%%%%%%%%%%%%%%%%%%%%%%%%%%%%%%%%%%%%%%%%%%%%%%%%%%%%%%%%%%%%%%%%%%%
%%%%%%%%%%%%%%%%%%%%%%%%%%%%%%%%%%%%%%%%%%%%%%%%%%%%%%%%%%%%%%%%%

\begin{table}[htb]
\caption{\label{Tab1}The values of the Hurst, multifractal scaling
 and generalized multifractal exponents for $q=2.0$,
for original, surrogate and shuffled of temperature fluctuation
series obtained by MF-DFA1.}
%\medskip
\begin{center}
\begin{tabular}{|c|c|c|c|}
%\hline
    Data & $H$ & $\tau$ &$D$ \\ \hline
   CMB  & $0.94\pm 0.01$ &$0.88\pm0.02$  &$0.88\pm0.02$     \\\hline
   Surrogate &$0.88\pm0.01$ &$0.76\pm0.02$ &$0.76 \pm 0.02$     \\\hline
  Shuffled & $0.50\pm0.001$ &$0.002\pm0.002$  & $0.002\pm0.002$  \\ %\hline
\end{tabular}
\end{center}
\end{table}

\section{Comparison of the multifractality for original,
shuffled and surrogate sunspot time series}

As discussed in the section III the remanning data set after the
elimination of sinusoidal trend has the multifractal nature. In
this section we are interested in to determine the source of
multifractality.
 In general, two different
types of multifractality in time series can be distinguished: (i)
Multifractality due to a fatness of probability density function
(PDF) of the time series. In this case the multifractality cannot
be removed by shuffling the series. (ii) Multifractality due to
different correlations in  small and large scale fluctuations. In
this case the data may have a PDF with finite moments, e.~g. a
Gaussian distribution. Thus the corresponding shuffled time series
will exhibit mono-fractal scaling, since all long-range
correlations are destroyed by the shuffling procedure. If both
kinds of multifractality are present, the shuffled series will
show weaker multifractality than the original series. The easiest
way to clarify the type of multifractality, is by analyzing the
corresponding shuffled and surrogate time series. The shuffling of
time series destroys the long range correlation, Therefore if the
multifractality only belongs to the long range correlation, we
should find $h_{\rm shuf}(q) = 0.5$. The multifractality nature
due to the fatness of the PDF signals is not affected by the
shuffling procedure. On the other hand, to determine the
multifractality due to the  broadness of PDF, the phase of
discrete fourier transform (DFT) coefficients of sunspot time
series are replaced with a set of pseudo independent distributed
uniform $(-\pi,\pi)$ quantities in the surrogate method. The
correlations in the surrogate series do not change, but the
probability function changes to the Gaussian distribution. If
multifractality in the time series is due to a broad PDF, $h(q)$
obtained by the surrogate method  will be independent of $q$. If
both kinds of multifractality are present in sunspot time series,
the shuffled and surrogate series will show weaker multifractality
than the original one.

\begin{figure}
\epsfxsize=8truecm\epsfbox{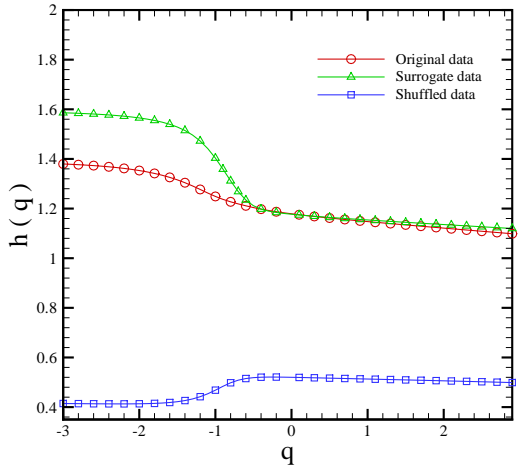} \narrowtext
\caption{Generalized Hurst exponent, $h(q)$ as a function of $q$
for original, surrogate and shuffled data.} \label{fig6}
 \end{figure}

To check the nature of multifractality,  we compare the
fluctuation function $F_q(s)$, for the original series ( after
cancelation of sinusoidal trend) with the result of the
corresponding shuffled, $F_q^{\rm shuf}(s)$ and surrogate series
$F_q^{\rm sur}(s)$. Differences between these two fluctuation
functions with the original one, directly indicate the presence of
long range correlations or broadness of probability density
function in the original series. These differences can be observed
in a plot of the ratio $F_q(s) / F_q^{\rm shuf}(s)$ and $F_q(s) /
F_q^{\rm sur}(s)$ versus $s$ \cite{bun02}. Since the anomalous
scaling due to a broad probability density affects $F_q(s)$ and
$F_q^{\rm shuf}(s)$ in the same way, only multifractality due to
correlations will be observed in $F_q(s) / F_q^{\rm shuf}(s)$. The
scaling behavior of these ratios are
\begin{equation} F_q(s) / F_q^{\rm shuf}(s) \sim s^{h(q)-h_{\rm
shuf}(q)} = s^{h_{\rm cor}(q)}, \label{HqCor} \end{equation}
\begin{equation} F_q(s) / F_q^{\rm sur}(s) \sim s^{h(q)-h_{\rm
sur}(q)} = s^{h_{\rm PDF}(q)}. \label{Hqpdf} \end{equation} If
only fatness of the PDF is responsible for the multifractality,
one should find $h(q)=h_{\rm shuf}(q)$ and $h_{\rm cor}(q)=0$. On
the other hand, deviations from $h_{\rm cor}(q) =0$ indicates the
presence of correlations, and $q$ dependence of $h_{\rm cor}(q)$
indicates that multifractality is due to the long rage
correlation. If only correlation multifractality is present, one
finds $h_{\rm shuf}(q)=0.5$. If both distribution and correlation
multifractality are present, both, $h_{\rm shuf}(q)$ and $h_{\rm
sur}(q)$ will depend on $q$. The $q$ dependence of the exponent
$h(q)$ for original, surrogate
 and shuffled time series are shown in
 Figures \ref{fig6}. The $q$ dependence of $h_{\rm cor}$ and
 $h_{\rm PDF}$ shows that the multifractality nature of sunspot time series is due
 to both broadness of the  PDF and long range correlation. The absolute value of $h_{\rm cor}(q)$ is
 greater than $h_{\rm PDF}(q)$, so the multifractality due to the fatness
 is weaker than the mulifractality due to the correlation.
 The deviation of $h_{\rm sur}(q)$ and $h_{\rm shuf}(q)$
 from $h(q)$ can be determined by using $\chi^2$ test as follows:
\begin{equation}
 \chi^2_{\diamond}=\sum_{i=1}^{N}\frac{[h(q_i)-h_{\diamond}(q_i)]^2}{\sigma(q_i)^2+\sigma_{\diamond}(q_i)^2},
 \label{khi} \end{equation}
the symbol $"\diamond"$ can be replaced by $"\rm sur"$ and $"\rm
shuf"$, to determine the confidence level of $h_{\rm sur}$ and
$h_{\rm shuf}$ to generalized Hurst exponents of original series,
respectively. The value of reduced chi-square
$\chi^2_{\nu\diamond}=\frac{\chi^2_{\diamond}}{\cal{N}}$
($\cal{N}$ is the number of degree of freedom) for shuffled and
surrogate time series are $1653.47$, $1.10$, respectively. On the
other hand the width of singularity spectrum, $f(\alpha)$, i.e.
$\Delta \alpha=\alpha(q_{min})-\alpha(q_{max})$ for original,
surrogate and shuffled time series are approximately, $0.44$,
$0.75$ and $0.22$ respectively. These values also show that the
multifractality due to correlation is dominant\cite{paw05}.

%%%%%%%%%%%%%%%%%%%%%%%%%%%%%%%%%%%%%%%%%%%%%%%%%%%%%%%%%%%%%%%%%%%%%%%%%
%%%%%%%%%%%%%%%%%%%%%%%%%%%%%%%%%%%%%%%%%%%%%%%%%%%%%%%%%%%%%%%%%

The values of the generalized Hurst exponent $h(q=2.0)$,
multifractal scaling $\tau(q=2)$ and generalized multifractal
exponents $(D(q=2))$ for the original, shuffled and surrogate of
sunspot time series obtained with MF-DFA1 method are reported in
Table \ref{Tab1}, The related scaling exponents are indicated in
Table \ref{Tab2}. The values of the Hurst exponent obtained by
MF-DFA1 and modified MF-DFA1 methods for original, surrogate and
shuffled sunspot time series are given in Table \ref{Tab3}.

\section{Conclusion}
The MF-DFA method allows us to determine the multifractal
characterization of the nonstationary and stationary time series.
The concept of MF-DFA of sunspot time series can be used to gain
deeper insight in to the processes occurring in nonstationary
dynamical system such as sunspots formation. We have shown that
the MF-DFA1 result of the monthly sunspot time series has three
crossover time scale $(s_{\times})$. These crossover time scale
are due to the sinusoidal trend. To minimizing the effect of this
trend, we have applied F-DFA on sunspot time series. Applying the
MF-DFA1 method on truncated data, demonstrated  that the monthly
sunspot time series is a nonstationary time series with
anti-correlation behavior. The $q$ dependence of $h(q)$ and
$\tau(q)$, indicated that the monthly sunspot time series has
multifractal behavior.  By comparing the generalized Hurst
exponent of the original time series with the shuffled and
surrogate one's, we have found that multifractality due to the
correlation has more contribution than the broadness of the
probability density function.
% Finally, the relative sunspot
%numbers, $R$ are a measure of solar activity on the entire disk of
%the sunspot \cite{tem02,bog82,ter02,ter03}. Here, according to the
%ani-correlation and multifractal behavior of sunspot time series,
%we concluded, the solar activity also has same statistical
%properties as sunspot time series.

\begin{table}
\begin{center}
\caption{\label{Tab1}The values of $h(q=2)$, multifractal scaling
and generalized multifractal exponents for $q=2.0$ for original,
surrogate and shuffled of monthly sunspot time series obtained by
MF-DFA1.}
%\medskip
\begin{tabular}{|c|c|c|c|}
 % \hline
    Data & $h$ & $\tau$ &$D$ \\ \hline
   Sunspot & $1.12\pm 0.01$ &$1.24\pm0.02$  &$1.24\pm0.02$     \\\hline
   Surrogate &$1.13\pm0.01$ &$1.26\pm0.02$ &$1.26\pm0.02$     \\\hline
  Shuffled & $0.51\pm0.01$ &$0.02\pm0.02$  & $0.02\pm0.02$  \\ %\hline
\end{tabular}
\end{center}
\end{table}
\begin{table}
\begin{center}
\caption{\label{Tab2}The values of the Hurst $(H)$, power spectrum
scaling $(\beta)$ and auto-correlation scaling $(\gamma)$
exponents for original, surrogate and shuffled of monthly sunspot
time series obtained by MF-DFA1.}
%\medskip
\begin{tabular}{|c|c|c|c|}
 % \hline
    Data & $H$ & $\beta$&$\gamma$\\ \hline
   Sunspot & $0.12\pm 0.02$ &$1.24\pm0.02$  &$-0.24\pm0.02$     \\\hline
   Surrogate &$0.13\pm0.02$ &$1.26\pm0.02$ &$-0.26\pm0.02$     \\\hline
  Shuffled & $0.51\pm0.01$ &$0.02\pm0.02$  & $0.98\pm0.02$  \\ %\hline
\end{tabular}
\end{center}
\end{table}
\begin{table}
\begin{center}
\caption{\label{Tab3}The value of the Hurst exponent using MF-DFA1
and modified MF-DFA1 for the original, shuffled and surrogate of
monthly sunspot time series.}
%\medskip
\begin{tabular}{|c|c|c|c|}
 % \hline
     Method & Sunspot & Surrogate &Shuffled \\ \hline
    MF-DFA1& $0.12\pm 0.01$ &$0.13\pm0.01$  &$0.51\pm0.01$     \\\hline
    Modified &$0.11\pm0.01$ &$0.12\pm0.01$ &$0.50\pm0.01$    \\
 \end{tabular}
\end{center}
\end{table}

{\bf Acknowledgements} We would like to thank Sepehr Arbabi
Bidgoli and Mojtaba Mohammadi Najafabadi for reading the
manuscript and useful comments. This paper is dedicated to Dr.
Somaihe Abdolahi.

\section{APPENDIX}
In this appendix we derive the relation between the exponent
$h(2)$ (DFA1 exponent) and Hurst exponent of a fBm signal. We show
that for such nonstationary signal the average sample variance
(Eq. \ref{fdef}) for $q=2$, is proportional to $s^{h(q)}$, where
$h(q=2)=H+1$. It is shown that the averaged sample variance
$F^2(s)$ behaves as:
\begin{eqnarray}
F^2(s)&\equiv&  \frac{1}{2N_s}\sum_{\nu=1}^{2N_s} \left
[F^2(s,\nu) \right],\nonumber\\
&=&\left\langle \left [F^2(s,\nu)
\right]\right \rangle_{\nu},\nonumber\\
&\equiv& {\mathcal{C}_{H}}s^{2(H+1)},\label{ap2311}
\end{eqnarray}
where $F^2(s,\nu)$ is defined as Eq. \ref{fsdef} and
${\mathcal{C}_{H}}$ is a function of Hurst exponent $H$.

To prove the statement we note that the data $x(k)$ is a
fractional Brownian motion (fBm), the partial sums $Y(i)$ (Eq.
\ref{profile}) will be a summed fBm signal.
 In the DFA1, the fitting function
will have the expression $(y_{\nu}=a_{\nu}+b_{\nu}i)$. The slope
$b_{\nu}$ and intercept $a_{\nu}$ of a least-squares line on
$Y(i)$ (from $0$ to $s$) for every windows $(\nu)$ are given by

\begin{eqnarray}
b_{\nu}&=&\frac{\sum_{i=1}^{s}Y(i)i-\frac{1}{s}\sum_{i=1}^{s}Y(i)\sum_{i=1}^{s}i
}{\sum_{i=1}^{s}i^2-\frac{1}{s}\left[\sum_{i=1}^{s}i\right]^2},\nonumber\\
&&\simeq \frac{\sum_{i=1}^{s}Y(i)i-\frac{s}{2}\sum_{i=1}^{s}Y(i)
}{s^3/12}\nonumber\\
a_{\nu}&=&\frac{1}{s}\sum_{i=1}^{s}Y(i)-\frac{1}{s}\sum_{i=1}^{s}i\simeq\frac{1}{s}\sum_{i=1}^{s}Y(i)-\frac{s}{2},
\label{ap2} \end{eqnarray} respectively.

 Using the Eqs. \ref{fdef} and \ref{ap2}, the Eq. \ref{ap2311} can be written as
 follows
\begin{eqnarray}
&&\left \langle\left [F^2(s,\nu) \right]\right\rangle
=\left\langle
\frac{1}{s}\sum_{i=1}^{s}(Y(i)-a-bi)^2\right \rangle\nonumber\\
&\simeq&\left\langle
\frac{1}{s}\sum_{i=1}^{s}Y(i)^2\right\rangle+\left\langle
a^2\right\rangle+\frac{s^2}{3}\left\langle
b^2\right\rangle\nonumber\\
&&-2\left\langle
\frac{a}{s}\sum_{i=1}^{s}y(i)\right\rangle-2\left\langle
\frac{b}{s}\sum_{i=1}^{s}iY(i)\right\rangle+s\left\langle ab
\right\rangle,\nonumber\\
&=&\left\langle\frac{1}{s}\sum_{i=1}^{s}Y(i)^2\right\rangle-\frac{4}{s^2}\left\langle\left[\sum_{i=1}^{s}Y(i)\right]^2\right\rangle\nonumber\\
&&-\frac{12}{s^4}\left\langle\left[\sum_{i=1}^{s}iY(i)\right]^2\right\rangle
+\frac{12}{s^3}\left\langle\sum_{i=1}^{s}iY(i)\sum_{i=1}^{s}Y(i)\right\rangle,\nonumber\\
&=&\frac{A}{s}-\frac{4}{s^2}B-\frac{12}{s^4}D+\frac{12}{s^3}C.
 \label{a3}
\end{eqnarray}
where we have discard the subscript $\nu$ for simplicity. Now let
us calculate the functions $A$, $B$, $C$ and $D$ in Eq. \ref{a3}.
 The increment of a summed fBm and fBm signals, i.e.
\begin{eqnarray}
x(i)=Y(i)-Y(i-1)\nonumber\\
u(i)=x(i)-x(i-1) \label{a4}
\end{eqnarray}
  are a fBm $x(i)$
and fGn $u(i)$ noise, respectively. The correlation of $Y(i)$ and
$x(i)$ are as follows \cite{murad}
\begin{eqnarray}
\left\langle x(i)x(j)\right\rangle&=&\frac{\sigma^2}{2}\left[i^{2H}+j^{2H}-|i-j|^{2H}\right],\nonumber\\
\left\langle
Y(i)Y(j)\right\rangle&=&\frac{\sigma^2}{(H+1)^2}\left(ij\right)^{H+1},
\label{a5}
\end{eqnarray}
where $\sigma^2=\left \langle u(i)^2\right\rangle$. Also the
variance of a summed fBm signal is $\left\langle
Y(i)^2\right\rangle=\frac{\sigma^2}{(H+1)^2}i^{2(H+1)}$
\cite{Peng94}. Finlly using the Eqs. \ref{ap2} and  \ref{a5}, it
can be easily shown that the Eq. \ref{a3} can be written as
follows
\begin{eqnarray}
\left \langle\left [F^2(s,\nu)
\right]\right\rangle_{\nu}={\mathcal{C}_{H}}s^{2(H+1)}, \label{a6}
\end{eqnarray}
where ${\mathcal{C}_{H}}$ is
\begin{eqnarray}
{\mathcal{C}_{H}}&=&\frac{\sigma^2}{(2H+3)(H+1)^2}-\frac{4\sigma^2}{[(H+1)(H+2)]^2}\nonumber\\
&&-\frac{12\sigma^2}{[(H+1)(H+3)]^2}+\frac{12\sigma^2}{(H+1)^2(H+2)(H+3)}.
\label{a7}
\end{eqnarray}

Therefore the standard DFA$1$ exponent for a nonstationary signal
is related to its Hurst exponent as $h(q=2)=H+1$.


\begin{thebibliography} {50}
\bibitem{note3}  For the maximum scale $s=N$ the fluctuation function
$F_q(s)$ is independent of $q$, since the sum in Eq.~(\ref{fdef})
runs over only two identical segments ($N_s \equiv [N/s] = 1$).
For smaller scales $s \ll N$ the averaging procedure runs over
several segments, and the average value $F_q(s)$ will be dominated
by the $F^2(s,\nu)$ from the segments with small (large)
fluctuations if $q < 0$ ($q > 0$).  Thus, for $s \ll N$, $F_q(s)$
with $q < 0$ will be smaller than $F_q(s)$ with $q > 0$, while
both become equal for $s=N$.  Hence, if we assume an homogeneous
scaling behavior of $F_q(s)$ following Eq.~(\ref{Hq}), the slope
$h(q)$ in a log-log plot of $F_q(s)$ with $q < 0$ versus $s$ must
be larger than the corresponding slope for $F_q(s)$ with $q > 0$.
Thus, $h(q)$ for $q < 0$ will usually be larger than $h(q)$ for $q
> 0$.
\end{thebibliography}
\end{document}